\documentclass[
 reprint,superscriptaddress,
 amsmath,amssymb,
 aps]{revtex4-1}

\usepackage{graphicx}
\usepackage{dcolumn}
\usepackage{bm}
\usepackage[ruled]{algorithm2e}
\usepackage{multirow,booktabs}
\usepackage{epstopdf}
\usepackage{svg}
\usepackage{paralist,siunitx,soul}
\usepackage[version=4]{mhchem}
\usepackage{hyperref}
\usepackage{url}

\hypersetup{colorlinks,urlcolor=blue,citecolor=blue,linkcolor=blue}
\DeclareUrlCommand\url{\color{blue}}

\graphicspath{{figs/}}

\begin{document}

\title{Bi-S network origin of cation-disorder stability and dispersive band edges in \ce{AgBiS2}}

\author{Han-Pu Liang}
\email{lianghanpu@nimte.ac.cn}
\affiliation{Advanced Interdisciplinary Science Research (AiR) Center, Ningbo Institute of Materials Technology and Engineering, Chinese Academy of Sciences, Ningbo 315201, China}

\author{Songyuan Geng}
\affiliation{Advanced Materials Thrust, Function Hub, The Hong Kong University of Science and Technology (Guangzhou), Guangzhou 511453, China}

\author{Heng Kang}
\affiliation{Advanced Interdisciplinary Science Research (AiR) Center, Ningbo Institute of Materials Technology and Engineering, Chinese Academy of Sciences, Ningbo 315201, China}

\author{Chen Qiu}
\affiliation{Department of Physics, Eastern Institute of Technology, Ningbo 315200, China}

\author{Xiao-Ping Yao}
\affiliation{Department of Physics, Eastern Institute of Technology, Ningbo 315200, China}

\author{Qing'an Li}
\affiliation{Beijing Computational Science Research Center, Beijing 100193, China}

\author{Bozhao Zhang}
\affiliation{Advanced Interdisciplinary Science Research (AiR) Center, Ningbo Institute of Materials Technology and Engineering, Chinese Academy of Sciences, Ningbo 315201, China}

\author{Lechuan Sun}
\affiliation{Advanced Interdisciplinary Science Research (AiR) Center, Ningbo Institute of Materials Technology and Engineering, Chinese Academy of Sciences, Ningbo 315201, China}

\author{Yuxuan Chen}
\affiliation{Advanced Interdisciplinary Science Research (AiR) Center, Ningbo Institute of Materials Technology and Engineering, Chinese Academy of Sciences, Ningbo 315201, China}

\author{Shan Zhang}
\affiliation{Advanced Interdisciplinary Science Research (AiR) Center, Ningbo Institute of Materials Technology and Engineering, Chinese Academy of Sciences, Ningbo 315201, China}

\author{Su-Huai Wei}
\affiliation{Department of Physics, Eastern Institute of Technology, Ningbo 315200, China}

\author{Peng-Fei Guan}
\email{pguan@nimte.ac.cn}
\affiliation{Advanced Interdisciplinary Science Research (AiR) Center, Ningbo Institute of Materials Technology and Engineering, Chinese Academy of Sciences, Ningbo 315201, China}


\begin{abstract}
	 Cation-disordered \ce{AgBiS2} is a promising lead-free optoelectronic material, but both its ordered structure and the microscopic origin of its favorable electronic properties remain debated. Theory has proposed a mixed-coordination tendency with tetrahedral \ce{AgS4} and octahedral \ce{BiS6} units, whereas experiments mainly report octahedrally coordinated ordered and cation-disordered phases, together with local cation off-centering. Here, we combine a machine-learning interatomic potential with a deep-learning Hamiltonian to resolve the coupled structural and electronic evolution of \ce{AgBiS2} at large length scales. We identify the three-dimensional Bi-S network as the central structural motif governing both disorder stability and band-edge electronic states. At weak disorder, Ag/Bi exchange competes with the off-centering tendency of the Ag sublattice, producing strongly distorted local environments and convoluted diffraction signatures that hinder the identification of the ordered phase. With increasing disorder, BiS$_6$-like units connect into a continuous Bi-S network, which stabilizes the rocksalt-like disordered phase. Despite strong cation disorder, \ce{AgBiS2} retains clear semiconductor-like band dispersion and develops a direct band gap. The connected Bi:$p$-S:$p$ states supported by the Bi-S network preserve a dispersive conduction-band edge and a small electron effective mass. In contrast, mobile Ag disrupts the long-range periodicity of Ag-S bonding, leading to strongly localized valence states. These results clarify the structural controversy in ordered \ce{AgBiS2} and establish a unified physical picture of disorder stability and optoelectronic response in nonisovalent semiconductor alloys.

\end{abstract}

\maketitle

\section{Introduction}

\ce{AgBiS2} has attracted broad interest as a lead-free semiconductor for photovoltaics, photodetection, and thermoelectrics \cite{2019-CM-AgBiS2, 2022-NP-AgBiS2, 2023-AM-AgBiS2, 2024-JACS-ABX2, 2025-ACSEL-AgBiS2,2025-ACSEL-AgBiS2-disorder, 2025-JACS-AgBiS2nano}. 
Its appeal comes from strong optical absorption, suitable band gaps, solution processability, and the use of relatively benign elements. A central structural puzzle, however, remains unresolved. Chemical-bonding arguments and first-principles calculations suggest that \ce{AgBiS2} should favor a mixed-coordination structure, in which Ag forms tetrahedral AgS$_4$ units and Bi forms octahedral BiS$_6$ units \cite{2024-JACS-ABX2}. In contrast, experiments usually In contrast, experiments usually identify low-temperature octahedrally coordinated ordered phases and high-temperature cation-disordered rocksalt-like phases \cite{2016-JMS-AgBiS}, while recent studies have further revealed local Ag/Bi off-centering within the ordered octahedral framework \cite{2026-arxiv-AgBiS2}. 
The absence of a clear mixed-coordination signature leaves the disordering pathway and structural stability of \ce{AgBiS2} unresolved. It is still unclear how mixed coordination competes with Ag/Bi chemical disorder during finite-temperature synthesis. This structural uncertainty further affects the interpretation of its optoelectronic properties, because local coordination and Ag/Bi ordering directly control the band edges, carrier transport, and optical response.

The unresolved optoelectronic mechanism of cation-disordered \ce{AgBiS2} reflects a broader challenge in disordered semiconductor alloys. Unlike ordered crystals, these materials retain only an average lattice framework, while atomic occupations are statistically disordered \cite{2018-NC-highentropy,2020-ACS-bandtail,2024-Nature-highentropy,2024-Nature-SQS}. The problem is especially severe in nonisovalent alloys, where local concentration fluctuations, antisite defects, and charge-imbalance motifs can introduce band-tail and defect-like localized states \cite{2002-Nature-bandgap,2014-PRB-ZnSnP2,2017-JPCC-ZnSnP2-expt-LRO,2019-PRM-CuGaSe2-disorder,2022-APLM-ZnGeN2,2024-PRB-AMX2}. These states blur the band edges and complicate the definition of the band gap \cite{2015-PRB-unconvergence,2022-ACSO-alloy-review,2026-SCPMA-dosf}. More importantly, they make it difficult to determine whether the electronic picture obtained from ordered phases can be transferred to the corresponding disordered phases. As a result, the optoelectronic properties of disordered alloys are often discussed in terms of qualitative trends, while the microscopic connection between local disorder, band-edge formation, and carrier transport remains unclear. Conventional density functional theory is usually too expensive for this task, particularly when finite-temperature disordering dynamics and momentum-resolved electronic structure need to be treated in the same framework.

Machine-learning methods provide a possible route to overcome this scale limitation. A machine-learning interatomic potential can sample long-time, large-scale molecular dynamics trajectories with near first-principles accuracy \cite{2024-SCM-nnap,2024-NM-nnap,2026-Acta-NNAP}. A deep-learning Hamiltonian can then predict electronic structures for large disordered snapshots without performing direct DFT calculations on each configuration \cite{2024-CPL-hamgnn,2024-NC-deeph,2025-npjcm-DLH,2026-Arxiv-deeph}. When combined with band unfolding, this strategy can recover momentum-resolved spectral functions and link atomic disorder to band-edge evolution \cite{2012-PRB-unfoldband,2020-JPCM-unfoldband}. However, such a unified approach has rarely been used to connect structural disorder, hidden phase signatures, and electronic properties in nonisovalent semiconductor alloys. \ce{AgBiS2} is therefore a useful model system. It combines an unresolved structure puzzle, strong cation disorder, and excellent optoelectronic performance, making it well suited for testing whether large-scale AI-accelerated simulations can reveal the coupled structural and electronic physics of disorder.

In this work, we combine a machine-learning interatomic potential with a deep-learning Hamiltonian method to study the structural and electronic evolution of cation-disordered \ce{AgBiS2} at large length scales. We identify the Bi-S network as the central motif linking cation disorder, structural stability, and band-edge electronic properties. In the weakly disordered regime, we examine the competition between Ag/Bi exchange, Ag-sublattice off-centering, and mixed coordination, which complicates the experimental identification of the ordered phase. In the strongly disordered regime, we establish the role of the three-dimensional Bi-S network in stabilizing the disordered structure. We further resolve the momentum-dependent electronic structure during disordering and distinguish the contrasting responses of the Ag-S-derived valence states and Bi-S-derived conduction states. This unified framework provides an atomistic and electronic description of cation-disordered \ce{AgBiS2} and clarifies the physical origin of its favorable optoelectronic properties.

\section{Methods}

\begin{figure}[!t]
	\includegraphics[width=8cm]{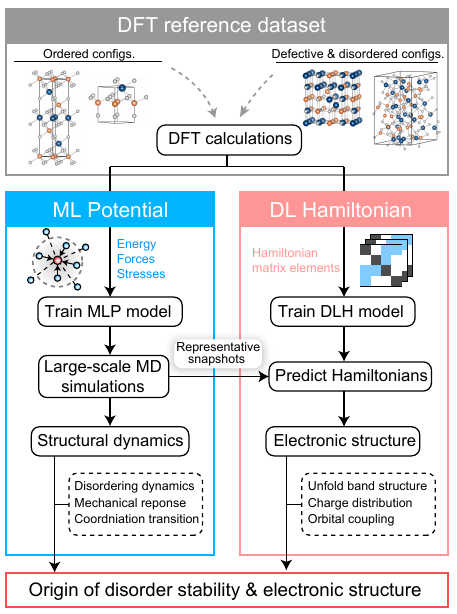}
	\caption{\label{fig:fig1} Workflow of Machine-learning potential (MLP) and Deep Learning Hamiltonian (DLH).}
\end{figure}

\begin{figure*}[!t]
	\includegraphics[width=16cm]{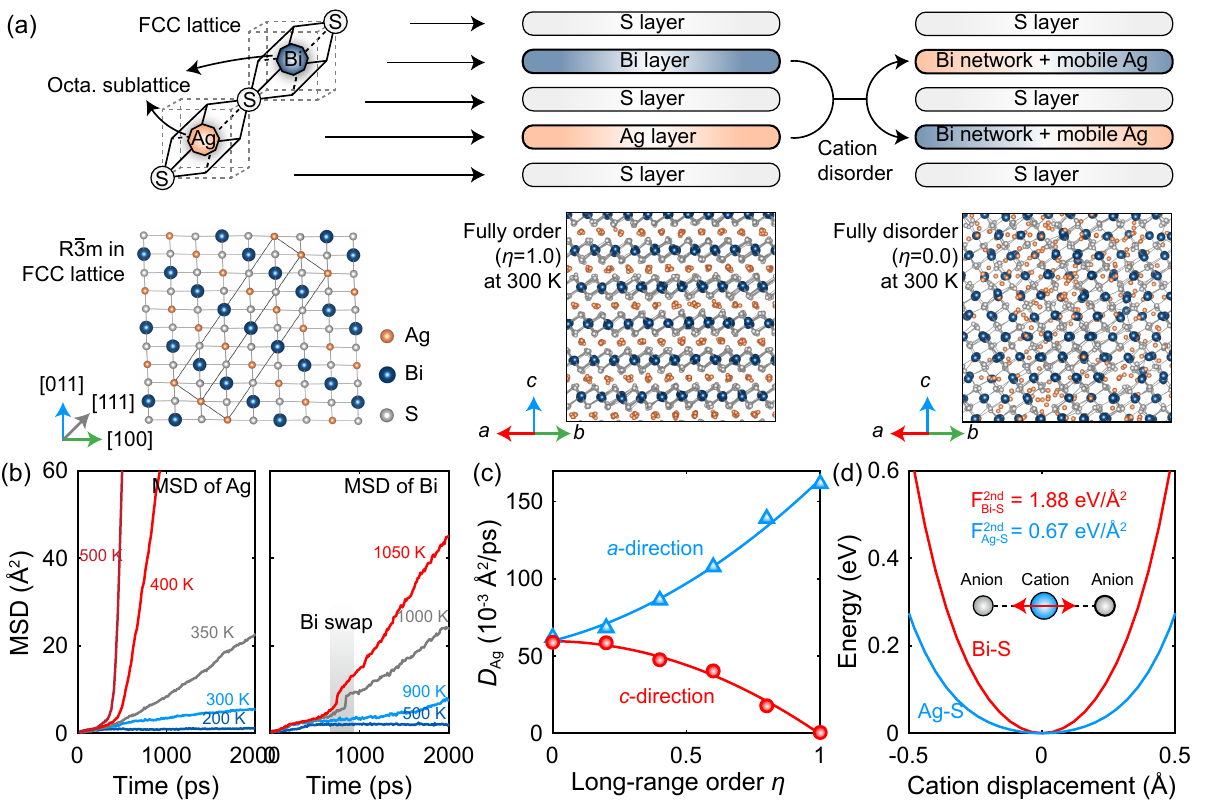}
	\caption{\label{fig:fig2} (a) Dynamic process of the order-disorder transition in \ce{AgBiS2}. (b) Mean square displacement (MSD) of Ag and (Bi as a function of time at different temperatures. (c) In-plane and out-of-plane diffusion coefficients of Ag as a function of the long-range order parameter. (d) Potential energy curves for cation displacement along the bond-stretching direction for Ag and Bi.}
\end{figure*}

To study the dynamical stability and electronic structure of large-scale disordered \ce{AgBiS2} at first-principles accuracy, we established a unified computational framework that combines a machine-learning interatomic potential (MLIP) with a deep-learning Hamiltonian (DLH), as shown in Figure \ref{fig:fig1}. The two parts of the framework are built on the same DFT dataset. On one side, the MLIP is trained to generate long-time, large-scale molecular dynamics trajectories, which are used to resolve the formation process and stability of the disordered structure. On the other side, the DLH model is trained to predict Hamiltonian matrices for representative disordered configurations extracted from the MLIP-driven trajectories. The predicted Hamiltonians are then used to obtain unfolded band structures, charge distributions, and orbital coupling information.

\textit{Dataset Contruction.} To cover the main local environments that may appear during the evolution of \ce{AgBiS2} from order to disorder, we constructed a DFT dataset including ordered, defective, and disordered structures. The ordered part contains the octahedrally coordinated structure $R\bar{3}m$ and the mixed-coordinated structure $P3m1$, which represent the ideal local bonding environments with different coordination preferences. The defective part includes vacancy defects and substitutional defects, which introduce local composition deviation and coordination distortion. The disordered part contains cation-disordered configurations with different long-range order parameters $\eta$. For each type of structure, NVT and NPT molecular dynamics sampling was performed over the temperature range from 50 to 1300 K, covering structural fluctuations from low-temperature vibration and thermally activated diffusion to the vicinity of melting. The final dataset contains 17,000 structures and serves as the dataset for both the MLIP and DLH training of disordered \ce{AgBiS2}. 

\textit{Machine-Learning Interatomic Potential.} The structural dynamics part is described by physics-informed neural
network atomic potential (NNAP) model \cite{2024-SCM-nnap,2024-NM-nnap,2026-Acta-NNAP}. This model uses a spherical harmonic-Chebyshev basis to represent the local atomic environment and encodes, within a cutoff radius, both the geometric distribution and the chemical species of neighboring atoms. The resulting local descriptors are then passed to a feedforward neural network to learn the mapping from atomic configuration to total energy. Atomic forces and stress tensors are obtained by automatic differentiation. For \ce{AgBiS2}, the NNAP model is trained simultaneously on energies, forces, and stresses. This design allows the model to describe not only equilibrium structures but also the diffusion, local deformation, and volume response involved in the disordering process. All structures in the dataset are calculated by VASP \cite{VASP-1,VASP-2,VASP-3,DFT-1, DFT-2} within PBE functional \cite{1996-PRL-PBE}, and the resulting energies, forces, and stresses are used as training labels for NNAP. After training, the mean absolute errors of energy, force, and stress are 1.38 meV/atom, 49.95 meV/\AA, and 0.803 meV/\AA$^3$, respectively.

\begin{figure*}[!t]
	\includegraphics[width=16cm]{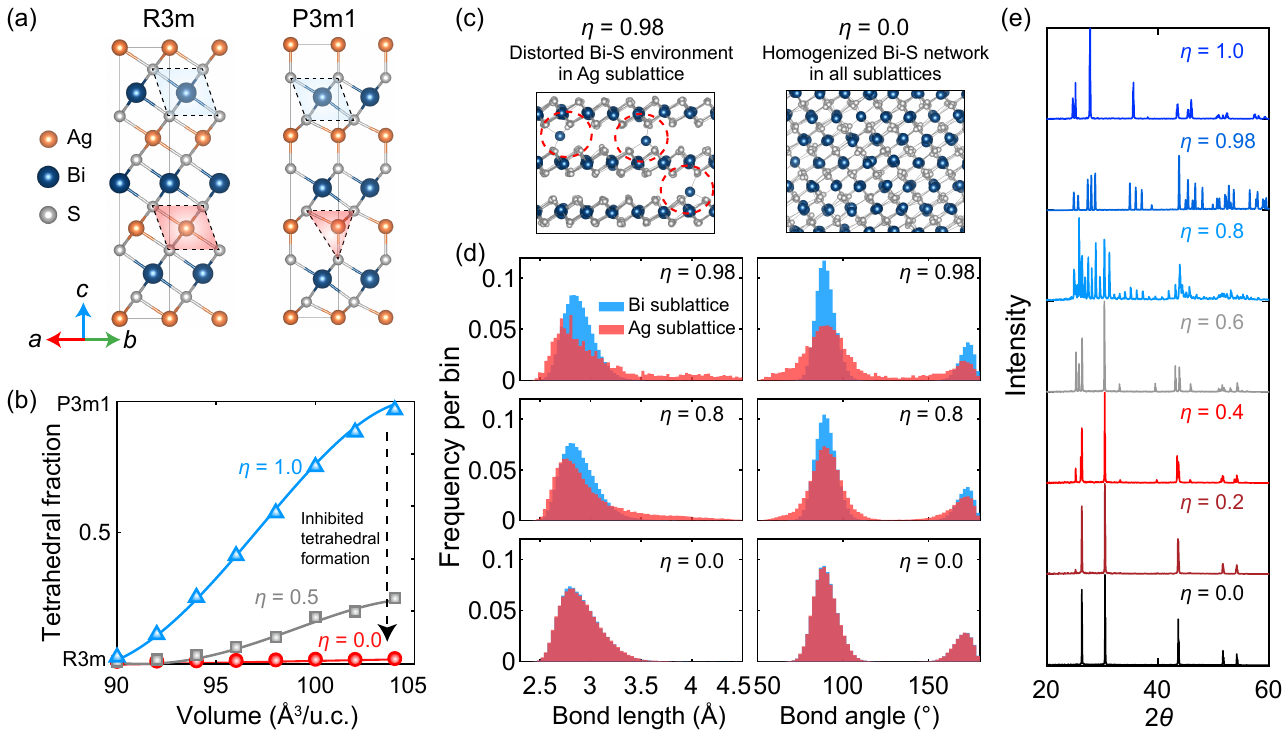}
	\caption{\label{fig:fig3} (a) Crystal structures of the octahedrally coordinated $R\bar{3}m$ and the mixed-coordination $P3m1$. (b) Tetrahedral fraction in the simulation supercell as a function of unit-cell volume at different long-range order parameters. (c) Snapshots of the Bi-S local environments at $\eta$=0.98 and 0.0. (d) Distributions of Bi-S bond lengths and S-Bi-S bond angles in Bi and Ag sublattices at $\eta$=0.98, 0.8 and 0.0. (e) Simulated XRD patterns of \ce{AgBiS2} at $\eta$=1.0, 0.8, 0.6, 0.4, 0.2 and 0.0. }
\end{figure*}

\textit{Deep-learning Hamiltonian.} The electronic structure part is described by the deep-learning Hamiltonian method \cite{2024-NC-deeph,2026-Arxiv-deeph,2024-CPL-hamgnn}. In this approach, real-space Hamiltonian matrix elements are predicted under a localized atomic-orbital basis, and the local atomic environment is encoded through a message-passing graph neural network. The network input depends only on relative coordinates and therefore naturally satisfies translational invariance. Rotational equivariance is preserved through equivariant features, spherical harmonics, and the corresponding tensor-coupling operations. Once the Hamiltonian is obtained, eigenvalues and eigenstates are calculated by diagonalization, and the band structure, spectral function, charge distribution, and orbital character of large-scale disordered structures are further derived with the QLDOS unfolding method \cite{2020-JPCM-unfoldband}. To cover different disordered local environments while controlling the training cost, representative configurations are selected in the crystal-feature vector space based on Mahalanobis distance. A total of 4,000 maximally separated configurations are retained, and their Hamiltonians are calculated by OPENMX \cite{2003-prb-openmx1,2004-prb-openmx2,2016-science-openmx3} within PBE functional as training labels. After training, the mean absolute error of the Hamiltonian matrix elements is about 1 meV.

Further details of the dataset composition, the training procedures, the hyperparameter settings, and the validation results of the NNAP and deep-learning Hamiltonian models are provided in the Supporting Information (SI). 

\section{Results and Discussion}

The octahedrally coordinated phase of ordered \ce{AgBiS2} adopts a hexagonal conventional cell with space group $R\bar{3}m$. It can be viewed as a face-centered-cubic S anion framework, in which Ag and Bi occupy octahedral cation sites and form a CuPt-type layered arrangement along the [111] direction, as shown in Figure \ref{fig:fig2}(a). In NPT molecular dynamics simulations accelerated by the MLIP, the ordered structure at 300 K retains the Ag/Bi layered feature, while Ag atoms already show clear off-center displacements. After heating to 1000 K and subsequent annealing to 300 K, Ag and Bi become statistically mixed on the cation sublattice. The ordered layered structure is then converted into a cation-disordered \ce{AgBiS2} alloy, where mobile Ag atoms coexist with a more rigid three-dimensional Bi-S network.

The disordering process is reflected in the mean square displacement of the cations. Figure \ref{fig:fig2}(b) shows that the MSD of Ag increases rapidly above 300 K, indicating that Ag diffusion is activated first. The origin lies in the fully occupied Ag:$d$-S:$p$ antibonding states near the valence-band maximum. These antibonding states flatten the local potential well around Ag and weaken its local coordination environment \cite{2024-PRB-AMX2}, thereby promoting off-center displacement and thermally activated migration. Upon further heating to about 1000 K, the MSD of Bi shows a distinct jump, indicating that part of the Bi atoms migrate from the Bi sublattice to the Ag sublattice through Ag/Bi exchange. This temperature range is consistent with the experimental synthesis process, where the disordered phase is usually obtained by high-temperature melting near 1087 K followed by annealing \cite{2016-JMS-AgBiS} . The agreement supports the reliability of the MLIP in describing the real cation-disordering dynamics.

The cation long-range order parameter is defined as $\eta = 2x - 1$ \cite{1992-PRL-LRO,2009-PRB-LRO}, where $x$ is the fraction of Bi atoms occupying the Bi sublattice. As $\eta$ decreases, the original Ag/Bi layered framework is progressively disrupted. Initially separated BiS$_6$ octahedra connect in three dimensions and form a percolating Bi-S network. At the same time, Ag diffusion changes from anisotropic layered transport to nearly isotropic three-dimensional transport, as illustrated in Figure \ref{fig:fig2}(c). The in-plane diffusion coefficient of Ag decreases, whereas the out-of-plane diffusion coefficient increases. In the fully disordered state, the two diffusion coefficients approach similar values. The microscopic driving force is the intrinsic bond strength difference between Ag-S and Bi-S. Figure \ref{fig:fig2}(d) shows that cation displacement energy curves along the bond stretching direction, indicating that the Bi-S bond has a much steeper potential energy profile than Ag-S, with a second-order force constant nearly three times larger. Once Bi enters the original Ag sublattice and becomes connected through S, these stronger Bi-S bonds form a rigid framework that supports the disordered phase and reshapes the Ag diffusion pathway.

\begin{figure*}[!t]
	\includegraphics[width=16cm]{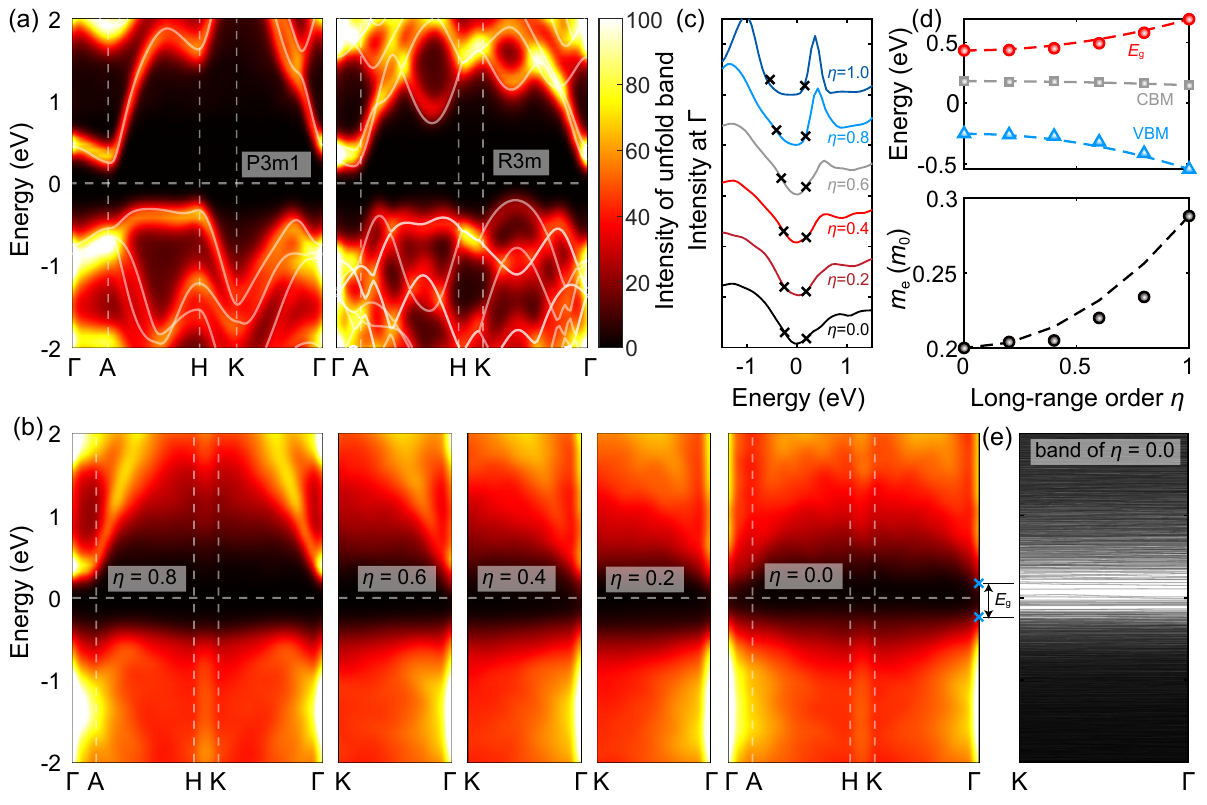}
	\caption{\label{fig:fig4} (a) Unfolded band structures of the ordered $P3m1$ and $R\bar{3}m$ phases at 300 K. (b) Unfolded band structures of disordered \ce{AgBiS2} with long-range order parameters $\eta$ = 0.8, 0.6, 0.4, 0.2, and 0.0. (c) Spectral functions at the $\Gamma$ point as a function of $\eta$. The black crosses mark the band-edge positions determined using the same spectral-intensity threshold. (d) Evolution of the band-edge positions, bandgap, and spectral electron effective mass as a function of $\eta$. (e) Folded band structure of the 1536-atom \ce{AgBiS2} supercell with $\eta = 0.0$.}
\end{figure*}

\ce{AgBiS2} favors a mixed-coordination structure in its ground state, where Ag forms tetrahedral AgS$_4$ units and Bi forms octahedral BiS$_6$ units, as shown in Figure \ref{fig:fig3}(a). 
Experiments, however, usually describe \ce{AgBiS2} in terms of octahedrally coordinated ordered and cation-disordered phases, although local Ag/Bi off-centering has also been reported \cite{2026-arxiv-AgBiS2}. A well-defined mixed-coordination phase with tetrahedral \ce{AgS4} and octahedral \ce{BiS6} units has nevertheless not been clearly established, suggesting that its structural signature may be masked when chemical disorder is introduced.

To examine how this motif evolves under disorder, we start from the $P3m1$ mixed-coordination structure and gradually introduce Ag/Bi exchange. The resulting structures show that the tetrahedral fraction is highly sensitive to both volume and cation order. In the ordered limit, the Ag sublattice can evolve between tetrahedral and octahedral coordination as the volume changes. As shown in Figure \ref{fig:fig3}(b), however, the tetrahedral fraction decreases rapidly once Ag/Bi disorder is introduced. This trend indicates that the mixed-coordination framework is fragile against cation exchange. The physical origin is that Bi atoms entering the Ag sublattice introduce stronger Bi-S bonds and resist the bond angle and bond length rearrangements required for a regular tetrahedral environment.

\begin{figure*}[!t]
	\includegraphics[width=16cm]{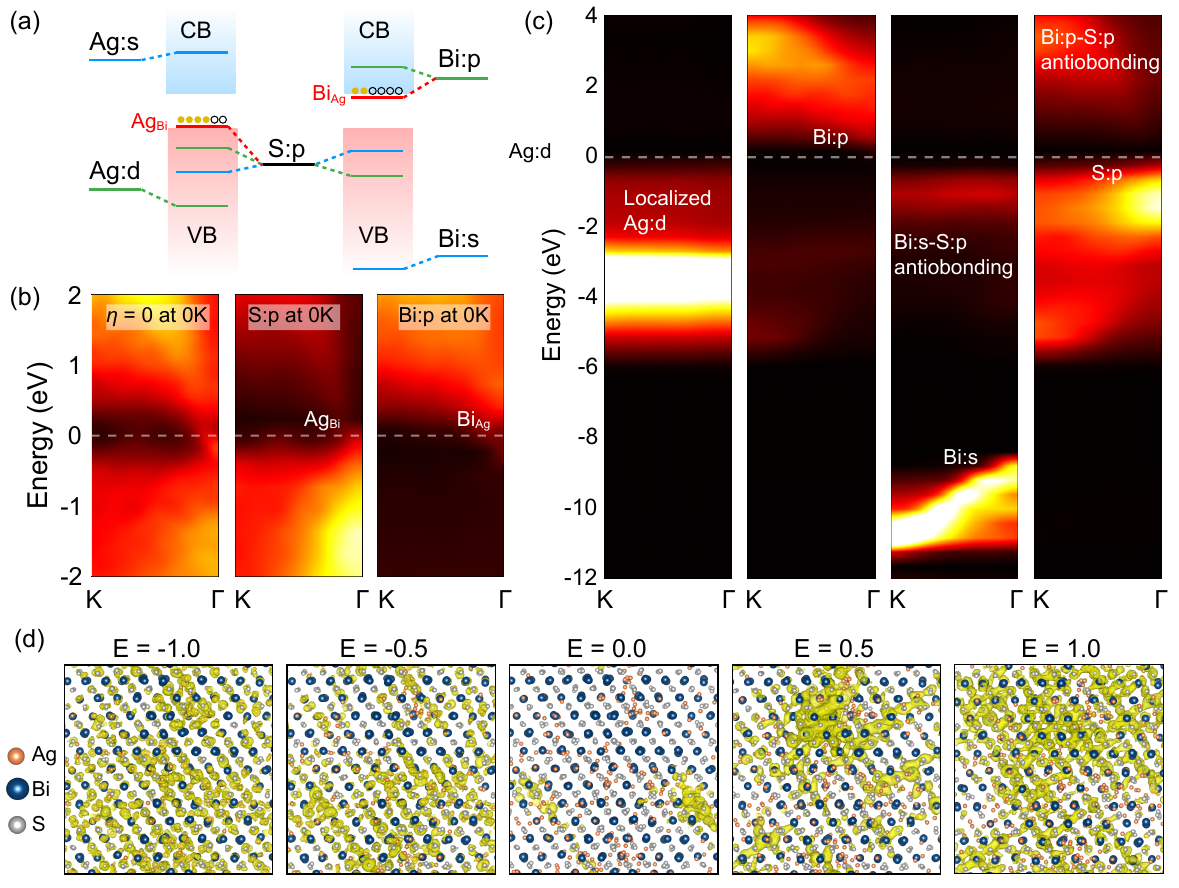}
	\caption{\label{fig:fig5} (a) Electronic orbital coupling diagram of ordered and antisite defected \ce{AgBiS2}.(b) Projected unfold band structure of total, S:$p$, and Bi:$p$ at 0 K. (c) Projected unfold band structure of Ag:$d$, Bi:$p$, Bi:$s$, and S:$p$. (d) Charge distribution at energy level of -1.0, -0.5, 0.0, 0.5, and 1.0 eV in unfold band structure of disordered \ce{AgBiS2}.}
\end{figure*}

The local structural response becomes evident even at very weak disorder. Figure \ref{fig:fig3}(c) compares the Bi-S environments at $\eta$ = 0.98 and $\eta$ = 0.0. At $\eta$ = 0.98, the Bi exchange ratio is only about 1\%, but the Bi atoms that enter the original Ag sublattice already show strong local distortion. This is because the Ag sublattice still largely preserves its tetrahedral character at high $\eta$, while Bi is unfavorable in such a tetrahedral sulfur environment according to the usual coordination preference implied by Pauling’s rule. As a result, Bi tends to recover BiS$_6$-like bonding locally, but the surrounding tetrahedral framework prevents the formation of a regular octahedron. The local Bi-S environment therefore becomes strongly distorted.

This picture is further supported by the Bi-S bond length and bond angle distributions in Figure \ref{fig:fig3}(d). At high $\eta$, Bi atoms on the Ag sublattice show much broader bond length and bond angle distributions than Bi atoms on the original Bi sublattice. As $\eta$ decreases, the difference between the two Bi environments becomes progressively smaller. In the fully disordered state, the Bi-S distributions on the two sublattices become nearly identical, indicating that Bi atoms now share a common octahedral-like environment within a three-dimensional Bi-S network.

These local distortions strongly affect the X-ray diffraction pattern. As shown in Figure \ref{fig:fig3}(e), at $\eta$ = 0.98, when a small amount of Bi is introduced into the Ag sublattice, the relaxed structures develop complex and disordered diffraction features. With further disordering, the pattern gradually evolves toward the cubic rocksalt-like diffraction profile. Since the XRD intensity of \ce{AgBiS2} is dominated by Bi, even a small number of Bi atoms in strongly distorted local environments can produce peak splitting, broadening, and apparent structural complexity. This provides a natural explanation for why low-temperature \ce{AgBiS2} samples often show poor crystallinity and complex diffraction features in experiments \cite{2010-ML-AgBiS2-xrd}. The mixed-coordination phase may therefore be hidden by weak Ag/Bi disorder rather than being completely absent.

To understand the origin of the favorable optoelectronic properties of cation-disordered \ce{AgBiS2}, we first examine the electronic structures of its ordered reference phases. 
The ordered $P3m1$ mixed-coordination phase and the ordered $R\bar{3}m$ octahedral phase differ in band dispersion and valence band maximum (VBM) position, but both are indirect bandgap semiconductors with the conduction band minimum (CBM) located at $\Gamma$, as shown in Figure \ref{fig:fig4}(a). Their bandgap sizes are also comparable. At the PBE level, both gaps are about 0.7 eV, while hybrid functional calculations give indirect bandgaps of 1.297 eV for $P3m1$ and 1.172 eV for $R\bar{3}m$.
The ideal ordered phases provide useful reference states, but they are insufficient to fully account for the optoelectronic behavior of \ce{AgBiS2}, especially when cation disorder is readily introduced under realistic conditions.

The unfolded band structures show a continuous evolution as $\eta$ decreases, as shown in Figure \ref{fig:fig4}(b). At weak disorder, the spectral features are already broadened, and the difference between the paths with different $k_z$ components, that is, $\Gamma\to A\to H$ and $K\to \Gamma$ is strongly reduced. This indicates that cation disorder disrupts the layer-stacking coherence inherited from the ordered structure, so that spectral weights from different $k_z$ paths begin to mix. With further disordering, the valence bands lose clear dispersion over most of the Brillouin zone. The spectral weight near the original off-$\Gamma$ VBM weakens, while the valence-band edge at $\Gamma$ shifts upward. As a result, \ce{AgBiS2} evolves from an indirect bandgap semiconductor to a direct bandgap semiconductor.

To track this evolution in a consistent way, we define the band edge at $\Gamma$ using a fixed spectral-intensity threshold. The resulting band gap decreases from about 0.7 eV in the ordered limit to about 0.5 eV in the fully disordered phase at the PBE level. The absolute gap values are expected to be underestimated, because PBE tends to over-delocalize electronic states and underestimate the band-edge separation \cite{2022-CPB-DFT}, but the disorder-dependent trend is meaningful. The gap follows an approximately $\eta^2$ dependence, as expected when the band-edge shift is mainly controlled by nearest-neighbor interactions in a disordered alloy \cite{2026-SCPMA-dosf}. 
The electron effective mass extracted from the dominant unfolded conduction branch also decreases with increasing disorder, as shown in Figure \ref{fig:fig4}(d), from about 0.3 $m_e$ in the ordered mixed-coordination phase to about 0.2 $m_e$ in the fully disordered phase. 
Because the fitting is performed on broadened spectral features, we refer to it as a spectral effective mass, with fitting details and uncertainty analysis provided in the SI. 
The decreasing electron effective mass suggests that cation disorder does not simply degrade transport. In \ce{AgBiS2}, disorder can also generate a more favorable conduction-band edge.

Furthermore, the need for an unfolded spectral description is directly illustrated by the band structure of the 1536-atom disordered supercell. As shown in Figure \ref{fig:fig4}(e), the bands contain many defective levels near the Fermi level. These states obscure the separation between occupied and unoccupied band edges, and the bandgap therefore cannot be reliably defined from the energy difference between occupied and unoccupied supercell eigenvalues. This does not mean that the material has metal characteristic. Instead, it is similar to the electronic structure of amorphous semiconductors, where localized tail states obscure the extended band edges \cite{2019-PSSA-amorphous,2022-amorphous-book}. The unfolded band structure resolves the dominant spectral features hidden behind these localized states and provides a more meaningful way to identify the band edges of disordered \ce{AgBiS2}. The size dependence of the electronic structure is discussed in detail in the SI.

The electronic structure of cation-disordered \ce{AgBiS2} is governed not only by chemical disorder but also by the accompanying structural relaxation. From the orbital-coupling picture shown in Figure \ref{fig:fig5}(a), the VBM of ordered \ce{AgBiS2} mainly originates from Ag:$d$-S:$p$ antibonding coupling, with dominant S:$p$ character. The CBM is primarily derived from Bi:$p$-S:$p$ coupling, with dominant Bi:$p$ character. Ag/Bi disorder introduces a high concentration of Ag$_{\rm Bi}$ and Bi$_{\rm Ag}$ antisite environments, which strongly perturb these orbital interactions.

In the fully octahedrally coordinated and chemically disordered structure before relaxation, the high concentrations of Ag$_{\rm Bi}$ and Bi$_{\rm Ag}$ antisites disrupt the local octet-rule valence matching around the S-centered coordination environments. The resulting local charge imbalance introduces many defect-like states near the Fermi level, as shown in Figure \ref{fig:fig5}(b). Without finite-temperature structural relaxation, the atomic framework cannot effectively accommodate the local valence mismatch, and these electronic states remain densely distributed around the Fermi level.

After relaxation at 300 K, the local structure and electronic states are substantially reorganized. Bi atoms preferentially recover octahedral coordination and connect through S atoms to form a three-dimensional Bi-S network. Meanwhile, the high mobility of Ag provides flexible local structural degrees of freedom that help accommodate the charge imbalance and restore local valence matching. As shown in Figure \ref{fig:fig5}(c), the relaxed structure consequently recovers distinguishable valence- and conduction-band regions. The Ag:$d$ states lose clear dispersion and become strongly localized because Ag diffusion destroys the periodicity of Ag-S bonding. By contrast, Bi:$p$ and S:$p$ states retain visible dispersion through the locally coordinated Bi-S framework. The lone-pair-related Bi:$s$-S:$p$ states, whose coupling is symmetry restricted at $\Gamma$ in the ordered structure, also lose their coherent dispersion as the BiS$_6$ units become spatially disordered.

The energy-resolved charge distributions provide a direct real-space view of this electronic reorganization. States near the Fermi level remain strongly localized. With increasing energy into the conduction band, the charge distribution progressively extends over a broader spatial range and is mainly concentrated on Bi and S atoms. On the valence-band side, the charge distribution also expands as the energy decreases, but it remains more localized and is mainly distributed over Ag and S atoms. This contrast explains why the conduction-band edge of disordered \ce{AgBiS2} retains a small effective mass. Although the BiS$_6$ octahedra are distributed randomly, the local Bi:$p$-S:$p$ bonding remains well defined, and the three-dimensional Bi-S network preserves long-range connectivity.

\section{Conclusion}

In this work, we established a unified large-scale computational framework that combines a machine-learning interatomic potential, a deep-learning Hamiltonian, and band unfolding analysis. 
Our results identify the Bi-S network as the central physical motif governing the evolution of \ce{AgBiS2} from ordered to disordered structures. 
In the weakly disordered regime, a small amount of Ag/Bi exchange competes with the off-centering tendency and mixed coordination of the Ag sublattice. 
The resulting local structural distortions generate convoluted diffraction features and hinder the identification of the ordered phase. 
This result provides a possible explanation for the long-standing discrepancy between theoretically predicted mixed coordination and experimentally reported octahedrally coordinated structures. 
As disorder increases, Bi atoms recover similar octahedral-like local environments on both cation sublattices. The resulting three-dimensional Bi-S network forms a rigid framework that stabilizes the disordered phase and changes Ag diffusion from layered transport toward nearly isotropic three-dimensional motion.
The same Bi-S network also governs the electronic structure of the disordered phase. Although cation disorder introduces many defect-like states and strongly broadens the valence bands, \ce{AgBiS2} retains semiconductor-like momentum-resolved dispersion and evolves toward a direct bandgap. The connected Bi:$p$-S:$p$ states supported by the Bi-S network preserve a dispersive conduction-band edge and a small spectral electron effective mass. By contrast, the high mobility of Ag disrupts the long-range periodicity of Ag-S bonding, causing Ag:$d$-derived valence states to become strongly localized.
The present framework provides a practical route for jointly resolving structural dynamics and electronic properties in large-scale disordered materials. More broadly, these results clarify the ordered-structure controversy in \ce{AgBiS2} and establish a transferable physical picture in which locally ordered bonding networks can stabilize strong chemical disorder while preserving favorable electronic transport channels.

\begin{acknowledgments}
	This work was supported by the Advanced Materials-National Science and Technology Major Project (Grant No. 2024ZD0606900), the National Natural Science Foundation of China (Grants No. T2325004, 12447158, 12504090, and 12504233), National Key Research and Development Program of China (Grant No. 2024YFA1409800), the Zhejiang Province National Science Foundation of China (Grant No. LQN26A040008, LQN26E010016, LQN26E010017), and the Talent Hub for "AI+ New Materials" Basic Research.
\end{acknowledgments}

\bibliography{my-references}

@article{2022-APLM-ZnGeN2,
  title = {Bandgap Analysis and Carrier Localization in Cation-Disordered {{ZnGeN2}}},
  author = {Cordell, Jacob J. and Tucker, Garritt J. and Tamboli, Adele and Lany, Stephan},
  year = {2022},
  month = jan,
  journal = {APL Mater.},
  volume = {10},
  number = {1},
  pages = {011112},
  issn = {2166-532X},
  doi = {10.1063/5.0077632},
  urldate = {2025-01-13},
  langid = {english},
}

@article{2002-Nature-bandgap,
  title = {Relationship between Local Structure and Phase Transitions of a Disordered Solid Solution},
  author = {Grinberg, Ilya and Cooper, Valentino R. and Rappe, Andrew M.},
  year = {2002},
  month = oct,
  journal = {Nature},
  volume = {419},
  number = {6910},
  pages = {909--911},
  issn = {0028-0836, 1476-4687},
  doi = {10.1038/nature01115},
  urldate = {2025-01-13},
  copyright = {http://www.springer.com/tdm},
  langid = {english},
}

@article{VASP-1,
  title = {Ab initio molecular dynamics for liquid metals},
  author = {Kresse, G. and Hafner, J.},
  journal = {Phys. Rev. B},
  volume = {47},
  issue = {1},
  pages = {558--561},
  numpages = {0},
  year = {1993},
  doi = {10.1103/PhysRevB.47.558}
}

@article{VASP-2,
  title = {Ab initio molecular-dynamics simulation of the liquid-metal amorphous-semiconductor transition in germanium},
  author = {Kresse, G. and Hafner, J.},
  journal = {Phys. Rev. B},
  volume = {49},
  issue = {20},
  pages = {14251--14269},
  numpages = {0},
  year = {1994},
  doi = {10.1103/PhysRevB.49.14251}
}

@article{VASP-3,
  title = {Efficient iterative schemes for ab initio total-energy calculations using a plane-wave basis set},
  author = {Kresse, G. and Furthm\"uller, J.},
  journal = {Phys. Rev. B},
  volume = {54},
  issue = {16},
  pages = {11169},
  numpages = {0},
  year = {1996},
  doi = {10.1103/PhysRevB.54.11169}
}

@article{DFT-1,
  title = {Inhomogeneous Electron Gas},
  author = {Hohenberg, P. and Kohn, W.},
  journal = {Phys. Rev.},
  volume = {136},
  issue = {3B},
  pages = {B864},
  numpages = {0},
  year = {1964},
  doi = {10.1103/PhysRev.136.B864}
}

@article{DFT-2,
  title = {Self-Consistent Equations Including Exchange and Correlation Effects},
  author = {Kohn, W. and Sham, L. J.},
  journal = {Phys. Rev.},
  volume = {140},
  issue = {4A},
  pages = {A1133},
  numpages = {0},
  year = {1965},
  doi = {10.1103/PhysRev.140.A1133}
}

@article{2009-PRB-LRO,
  title = {Comparison of Atomistic Simulations and Statistical Theories for Variable Degree of Long-Range Order in Semiconductor Alloys},
  author = {Zhang, Yong and Mascarenhas, A. and Wei, Su-Huai and Wang, L.-W.},
  year = {2009},
  month = jul,
  journal = {Phys. Rev. B},
  volume = {80},
  number = {4},
  pages = {045206},
  issn = {1098-0121, 1550-235X},
  doi = {10.1103/PhysRevB.80.045206},
  urldate = {2024-12-29},
  copyright = {http://link.aps.org/licenses/aps-default-license},
  langid = {english},
}

@article{2014-PRB-ZnSnP2,
  title = {Origin of the Failed Ensemble Average Rule for the Band Gaps of Disordered Nonisovalent Semiconductor Alloys},
  author = {Ma, Jie and Deng, Hui-Xiong and Luo, Jun-Wei and Wei, Su-Huai},
  year = {2014},
  month = sep,
  journal = {Phys. Rev. B},
  volume = {90},
  number = {11},
  pages = {115201},
  issn = {1098-0121, 1550-235X},
  doi = {10.1103/PhysRevB.90.115201},
  urldate = {2024-12-26},
  copyright = {http://link.aps.org/licenses/aps-default-license},
  langid = {english},
}

@article{2015-PRB-unconvergence,
  title = {Special Quasirandom Structure in Heterovalent Ionic Systems},
  author = {Seko, Atsuto and Tanaka, Isao},
  year = {2015},
  month = jan,
  journal = {Phys. Rev. B},
  volume = {91},
  number = {2},
  pages = {024106},
  issn = {1098-0121, 1550-235X},
  doi = {10.1103/PhysRevB.91.024106},
  urldate = {2024-12-29},
  copyright = {http://link.aps.org/licenses/aps-default-license},
  langid = {english},
}

@article{2023-AM-AgBiS2,
  title = {Cation-{{Disorder Engineering Promotes Efficient Charge}}-{{Carrier Transport}} in {{AgBiS}}{\textsubscript{2}} {{Nanocrystal Films}}},
  author = {Righetto, Marcello and Wang, Yongjie and Elmestekawy, Karim A. and Xia, Chelsea Q. and Johnston, Michael B. and Konstantatos, Gerasimos and Herz, Laura M.},
  year = {2023},
  month = nov,
  journal = {Adv. Mater.},
  volume = {35},
  number = {48},
  pages = {2305009},
  issn = {0935-9648, 1521-4095},
  doi = {10.1002/adma.202305009},
  urldate = {2024-12-29},
  langid = {english},
}

@article{2017-JPCC-ZnSnP2-expt-LRO,
  title = {Order--{{Disorder Phenomena}} and {{Their Effects}} on {{Bandgap}} in {{ZnSnP}}{\textsubscript{2}}},
  author = {Nakatsuka, Shigeru and Nose, Yoshitaro},
  year = {2017},
  month = jan,
  journal = {J. Phys. Chem. C},
  volume = {121},
  number = {2},
  pages = {1040--1046},
  issn = {1932-7447, 1932-7455},
  doi = {10.1021/acs.jpcc.6b11215},
  urldate = {2024-12-28},
  langid = {english},
}

@article{1992-PRL-LRO,
  title = {Evolution of Alloy Properties with Long-Range Order},
  author = {Laks, David B. and Wei, Su-Huai and Zunger, Alex},
  year = {1992},
  month = dec,
  journal = {Phys. Rev. Lett.},
  volume = {69},
  number = {26},
  pages = {3766--3769},
  issn = {0031-9007},
  doi = {10.1103/PhysRevLett.69.3766},
  urldate = {2024-12-26},
  copyright = {http://link.aps.org/licenses/aps-default-license},
  langid = {english}
}

@article{2019-PRM-CuGaSe2-disorder,
  title = {Predicting Copper Gallium Diselenide and Band Structure Engineering through Order-Disordered Transition},
  author = {Liu, Wenjie and Liang, Hanpu and Duan, Yifeng and Wu, Zhigang},
  year = {2019},
  month = dec,
  journal = {Phys. Rev. Mater.},
  volume = {3},
  number = {12},
  pages = {125405},
  issn = {2475-9953},
  doi = {10.1103/PhysRevMaterials.3.125405},
  urldate = {2024-12-28},
  copyright = {CC0 1.0 Universal Public Domain Dedication},
  langid = {english},
}

@article{2022-ACSO-alloy-review,
  title = {Energy {{Scales}} of {{Compositional Disorder}} in {{Alloy Semiconductors}}},
  author = {Baranovskii, Sergei D. and Nenashev, Alexey V. and Hertel, Dirk and Gebhard, Florian and Meerholz, Klaus},
  year = {2022},
  month = dec,
  journal = {ACS Omega},
  volume = {7},
  number = {50},
  pages = {45741--45751},
  issn = {2470-1343, 2470-1343},
  doi = {10.1021/acsomega.2c05426},
  urldate = {2024-12-28},
  copyright = {https://creativecommons.org/licenses/by-nc-nd/4.0/},
  langid = {english},
}

@article{2024-PRB-AMX2,
  title = {Unveiling Disparities and Promises of {{Cu}} and {{Ag}} Chalcopyrites for Thermoelectrics},
  author = {Liang, Han-Pu and Geng, Songyuan and Jia, Tiantian and Li, Chuan-Nan and Xu, Xun and Zhang, Xie and Wei, Su-Huai},
  year = {2024},
  month = jan,
  journal = {Phys. Rev. B},
  volume = {109},
  number = {3},
  pages = {035205},
  issn = {2469-9950, 2469-9969},
  doi = {10.1103/PhysRevB.109.035205},
  urldate = {2024-05-07},
  langid = {english},
}

@article{2024-JACS-ABX2,
  title = {Critical {{Role}} of {{Configurational Disorder}} in {{Stabilizing Chemically Unfavorable Coordination}} in {{Complex Compounds}}},
  author = {Liang, Han-Pu and Li, Chuan-Nan and Zhou, Ran and Xu, Xun and Zhang, Xie and Yang, Jingxiu and Wei, Su-Huai},
  year = {2024},
  month = jun,
  journal = {J. Am. Chem. Soc.},
  volume = {146},
  number = {23},
  pages = {16222--16228},
  issn = {0002-7863, 1520-5126},
  doi = {10.1021/jacs.4c04201},
  urldate = {2024-12-28},
  copyright = {https://doi.org/10.15223/policy-029},
  langid = {english},
}

@article{2022-NP-AgBiS2,
  title = {Cation Disorder Engineering Yields {{AgBiS2}} Nanocrystals with Enhanced Optical Absorption for Efficient Ultrathin Solar Cells},
  author = {Wang, Yongjie and Kavanagh, Se{\'a}n R. and {Burgu{\'e}s-Ceballos}, Ignasi and Walsh, Aron and Scanlon, David O. and Konstantatos, Gerasimos},
  year = {2022},
  month = mar,
  journal = {Nat. Photon.},
  volume = {16},
  number = {3},
  pages = {235--241},
  issn = {1749-4885, 1749-4893},
  doi = {10.1038/s41566-021-00950-4},
  urldate = {2024-12-28},
  langid = {english},
}

@article{2019-CM-AgBiS2,
author = {Rathore, Ekashmi and Juneja, Rinkle and Culver, Sean P. and Minafra, Nicolò and Singh, Abhishek K. and Zeier, Wolfgang G. and Biswas, Kanishka},
title = {Origin of Ultralow Thermal Conductivity in n-Type Cubic Bulk \ce{AgBiS2}: Soft Ag Vibrations and Local Structural Distortion Induced by the Bi 6s2 Lone Pair},
journal = {Chem. Mater.},
volume = {31},
number = {6},
pages = {2106-2113},
year = {2019},
doi = {10.1021/acs.chemmater.9b00001}
}

@article{2016-JMS-AgBiS,
    title = { Thermochemical properties of selected ternary phases in the Ag-Bi-S system },
    author = { Tesfaye, Fiseha and Lindberg, Daniel },
    journal = { J. Mater. Sci. },
    volume = { 51 },
    number = { 12 },
    pages = { 5750 },
    year = { 2016 },
    doi = { 10.1007/s10853-016-9877-8 },
    url = { https://doi.org/10.1007/s10853-016-9877-8 },
}

@article{2025-ACSEL-AgBiS2,
author = {Yang, Wanpeng and Sun, Tianyu and Ma, Xiaoting and Yu, Haixuan and Shi, Haodan and Hu, Yong and Huang, Junyi and Liu, Zhirong and Xu, Ying and Li, Xiongjie and Shen, Yan and Wang, Mingkui},
title = {Boosting Open-Circuit Voltage of AgBiS2 Quantum Dot Solar Cells through Post-treatment Passivation},
journal = {ACS Energy Lett.},
volume = {10},
number = {1},
pages = {58-67},
year = {2025},
doi = {10.1021/acsenergylett.4c03015},
}

@article{2025-JACS-AgBiS2nano,
    title = { Triplet Sensitization Photon Upconversion Using Near-Infrared Indirect-Bandgap AgBiS2 Nanocrystals },
    author = { Chang, Kin Ting and Liang, Wenfei and Gong, Shaokuan and Yeung, Pang Ho and Feng, Jianning and Chen, Xihan and Lu, Haipeng },
    journal = { J. Am. Chem. Soc. },
    volume = { 147 },
    number = { 16 },
    pages = { 14015 },
    year = { 2025 },
    doi = { 10.1021/jacs.5c04015 },
    url = { https://doi.org/10.1021/jacs.5c04015 },
}

@article{2025-ACSEL-AgBiS2-disorder,
    title = { Vacancy-Catalyzed Cation Homogenization for High-Performance AgBiS2 Nanocrystal Solar Cells },
    author = { Liu, Yang and Ni, Zitao and Peng, Lucheng and Wu, Hao and Liu, Zeke and Wang, Yongjie and Ma, Wanli and Konstantatos, Gerasimos },
    journal = { ACS Energy Lett. },
    volume = { 10 },
    number = { 4 },
    pages = { 2068 },
    year = { 2025 },
    doi = { 10.1021/acsenergylett.5c00506 },
    url = { https://doi.org/10.1021/acsenergylett.5c00506 },
}

@article{2018-NC-highentropy,
  title = {High-Entropy High-Hardness Metal Carbides Discovered by Entropy Descriptors},
  author = {Sarker, Pranab and Harrington, Tyler and Toher, Cormac and Oses, Corey and Samiee, Mojtaba and Maria, Jon-Paul and Brenner, Donald W. and Vecchio, Kenneth S. and Curtarolo, Stefano},
  year = {2018},
  month = nov,
  journal = {Nat. Commun.},
  volume = {9},
  number = {1},
  pages = {4980},
  issn = {2041-1723},
  doi = {10.1038/s41467-018-07160-7},
  urldate = {2024-12-29},
  langid = {english},
}

@article{2020-ACS-bandtail,
  title = {Band {{Tails}} and {{Cu}}--{{Zn Disorder}} in {{Cu2ZnSnS4 Solar Cells}}},
  author = {Larsen, J. K. and Scragg, J. J. S. and Ross, N. and {Platzer-Bj{\"o}rkman}, C.},
  year = {2020},
  month = aug,
  journal = {ACS Appl. Energy Mater.},
  volume = {3},
  number = {8},
  pages = {7520--7526},
  publisher = {American Chemical Society},
  doi = {10.1021/acsaem.0c00926},
}

@article{2024-Nature-highentropy,
  title = {Disordered Enthalpy--Entropy Descriptor for High-Entropy Ceramics Discovery},
  author = {Divilov, Simon and Eckert, Hagen and Hicks, David and Oses, Corey and Toher, Cormac and Friedrich, Rico and Esters, Marco and Mehl, Michael J. and Zettel, Adam C. and Lederer, Yoav and Zurek, Eva and Maria, Jon-Paul and Brenner, Donald W. and Campilongo, Xiomara and Filipovi{\'c}, Suzana and Fahrenholtz, William G. and Ryan, Caillin J. and DeSalle, Christopher M. and Crealese, Ryan J. and Wolfe, Douglas E. and Calzolari, Arrigo and Curtarolo, Stefano},
  year = {2024},
  month = jan,
  journal = {Nature},
  volume = {625},
  number = {7993},
  pages = {66--73},
  issn = {0028-0836, 1476-4687},
  doi = {10.1038/s41586-023-06786-y},
  urldate = {2024-12-29},
  langid = {english},
}

@article{2024-Nature-SQS,
  title = {Chemical Short-Range Disorder in Lithium Oxide Cathodes},
  author = {Wang, Qidi and Yao, Zhenpeng and Wang, Jianlin and Guo, Hao and Li, Chao and Zhou, Dong and Bai, Xuedong and Li, Hong and Li, Baohua and Wagemaker, Marnix and Zhao, Chenglong},
  year = {2024},
  month = may,
  journal = {Nature},
  volume = {629},
  number = {8011},
  pages = {341--347},
  issn = {0028-0836, 1476-4687},
  doi = {10.1038/s41586-024-07362-8},
  urldate = {2025-01-11},
  langid = {english},
}

@article{2026-SCPMA-dosf,
  author = "Liang, Han-Pu and Li, Chuan-Nan and Tang, Xin-Ru and Xu, Xun and Qiu, Chen and Huang, Qiu-Shi and Wei, Su-Huai",
  title = "Effect of concentration fluctuations on material properties of disordered alloys",
  journal = "Sci. China-Phys. Mech. Astron.",
  year = "2026",
  volume = "69",
  number = "4",
  pages = "247311",
  doi = "https://doi.org/10.1007/s11433-025-2885-3"
}

@article{2026-Acta-NNAP,
title = {Atomic-bond-modulated thermal transport in Si/AlN heterointerfaces revealed via machine-learning potentials},
journal = {Acta Mater.},
volume = {312},
pages = {122241},
year = {2026},
issn = {1359-6454},
doi = {https://doi.org/10.1016/j.actamat.2026.122241},
url = {https://www.sciencedirect.com/science/article/pii/S1359645426003459},
author = {Yuxuan Chen and Qing-an Li and Hanpu Liang and Heng Kang and Guanchen Dong and Kexin Zhang and Lin Wang and Huanrong Liu and Shan Zhang and Jian Li and Bin Xu and Xuelin Yang and Shiwu Gao and Rui Su and Pengfei Guan},
}

@article{2024-SCM-nnap,
    title = { Efficient and accurate simulation of vitrification in multicomponent metallic liquids with neural network potentials },
    author = { Su, Rui and Yu, Jieyi and Guan, Pengfei and Wang, Weihua },
    journal = { Sci. China Mater. },
    volume = { 67 },
    number = { 10 },
    pages = { 3298 },
    year = { 2024 },
    doi = { 10.1007/s40843-024-2953-9 },
    url = { https://doi.org/10.1007/s40843-024-2953-9 },
}

@article{2024-NM-nnap,
    title = { Oxidation-induced superelasticity in metallic glass nanotubes },
    author = { Li, Fucheng and Zhang, Zhibo and Liu, Huanrong and Zhu, Wenqing and Wang, Tianyu and Park, Minhyuk and Zhang, Jingyang and B\:onninghoff, Niklas and Feng, Xiaobin and Zhang, Hongti and Luan, Junhua and Wang, Jianguo and Liu, Xiaodi and Chang, Tinghao and Chu, Jinn P. and Lu, Yang and Liu, Yanhui and Guan, Pengfei and Yang, Yong },
    journal = { Nat. Mater.},
    volume = { 23 },
    number = { 1 },
    pages = { 52 },
    year = { 2024 },
    doi = { 10.1038/s41563-023-01733-8 },
    url = { https://doi.org/10.1038/s41563-023-01733-8 },
}

@article{2025-npjcm-DLH,
    title = { Active learning of effective Hamiltonian for super-large-scale atomic structures },
    author = { Ma, Xingyue and Chen, Hongying and He, Ri and Yu, Zhanbo and Prokhorenko, Sergei and Wen, Zheng and Zhong, Zhicheng and \'{I}\~{n}iguez-Gonz\'{a}lez, Jorge and Bellaiche, L. and Wu, Di and Yang, Yurong },
    journal = { npj Comput. Mater. },
    volume = { 11 },
    number = { 1 },
    pages = { 70 },
    year = { 2025 },
    doi = { 10.1038/s41524-025-01563-z },
    url = { https://doi.org/10.1038/s41524-025-01563-z },
}

@article{2024-CPL-hamgnn,
doi = {10.1088/0256-307X/41/7/077103},
url = {https://doi.org/10.1088/0256-307X/41/7/077103},
year = {2024},
month = {jun},
publisher = {Chinese Physical Society and IOP Publishing Ltd},
volume = {41},
number = {7},
pages = {077103},
author = {Zhong, Yang and Yu, Hongyu and Yang, Jihui and Guo, Xingyu and Xiang, Hongjun and Gong, Xingao},
title = {Universal Machine Learning Kohn-Sham Hamiltonian for Materials},
journal = {Chin. Phys. Lett.},
}

@article{2026-Arxiv-deeph,
    title={DeepH-pack: A general-purpose neural network package for deep-learning electronic structure calculations},
    author={Li, Yang and Wang, Yanzhen and Zhao, Boheng and Gong, Xiaoxun and Wang, Yuxiang and Tang, Zechen and Wang, Zixu and Yuan, Zilong and Li, Jialin and Sun, Minghui and Chen, Zezhou and Tao, Honggeng and Wu, Baochun and Yu, Yuhang and Li, He and da Jornada, Felipe H. and Duan, Wenhui and Xu, Yong },
    journal={arXiv preprint arXiv:2601.02938},
    year={2026}
}

@article{2024-NC-deeph,
    title = { A deep equivariant neural network approach for efficient hybrid density functional calculations },
    author = { Tang, Zechen and Li, He and Lin, Peize and Gong, Xiaoxun and Jin, Gan and He, Lixin and Jiang, Hong and Ren, Xinguo and Duan, Wenhui and Xu, Yong },
    journal = { Nat. Commun. },
    volume = { 15 },
    number = { 1 },
    pages = { 8815 },
    year = { 2024 },
    doi = { 10.1038/s41467-024-53028-4 },
    url = { https://doi.org/10.1038/s41467-024-53028-4 },
}

@article{2012-PRB-unfoldband,
  title = {Extracting $E$ versus $\stackrel{P\vec}{k}$ effective band structure from supercell calculations on alloys and impurities},
  author = {Popescu, Voicu and Zunger, Alex},
  journal = {Phys. Rev. B},
  volume = {85},
  issue = {8},
  pages = {085201},
  numpages = {12},
  year = {2012},
  month = {Feb},
  publisher = {American Physical Society},
  doi = {10.1103/PhysRevB.85.085201},
  url = {https://link.aps.org/doi/10.1103/PhysRevB.85.085201}
}

@article{2020-JPCM-unfoldband,
doi = {10.1088/1361-648X/ab6e8e},
url = {https://doi.org/10.1088/1361-648X/ab6e8e},
year = {2020},
month = {feb},
publisher = {IOP Publishing},
volume = {32},
number = {20},
pages = {205902},
author = {Mayo, Sara G and Yndurain, Felix and Soler, Jose M},
title = {Band unfolding made simple},
journal = {J. Phys.: Condens. Matter},
}

@article{1996-PRL-PBE,
  title = {Generalized Gradient Approximation Made Simple},
  author = {Perdew, John P. and Burke, Kieron and Ernzerhof, Matthias},
  journal = {Phys. Rev. Lett.},
  volume = {77},
  issue = {18},
  pages = {3865--3868},
  numpages = {0},
  year = {1996},
  month = {Oct},
  publisher = {American Physical Society},
  doi = {10.1103/PhysRevLett.77.3865},
  url = {https://link.aps.org/doi/10.1103/PhysRevLett.77.3865}
}

@article{2003-prb-openmx1,
  title = {Variationally optimized atomic orbitals for large-scale electronic structures},
  author = {Ozaki, T.},
  journal = {Phys. Rev. B},
  volume = {67},
  issue = {15},
  pages = {155108},
  numpages = {5},
  year = {2003},
  month = {Apr},
  publisher = {American Physical Society},
  doi = {10.1103/PhysRevB.67.155108},
  url = {https://link.aps.org/doi/10.1103/PhysRevB.67.155108}
}

@article{2004-prb-openmx2,
  title = {Numerical atomic basis orbitals from H to Kr},
  author = {Ozaki, T. and Kino, H.},
  journal = {Phys. Rev. B},
  volume = {69},
  issue = {19},
  pages = {195113},
  numpages = {19},
  year = {2004},
  month = {May},
  publisher = {American Physical Society},
  doi = {10.1103/PhysRevB.69.195113},
  url = {https://link.aps.org/doi/10.1103/PhysRevB.69.195113}
}

@article{2016-science-openmx3,
author = {Kurt Lejaeghere and Gustav Bihlmayer and Torbj\"{o}rn Bj\"{o}rkman and others},
title = {Reproducibility in density functional theory calculations of solids},
journal = {Science},
volume = {351},
number = {6280},
pages = {aad3000},
year = {2016},
doi = {10.1126/science.aad3000},
}

@misc{2026-arxiv-AgBiS2,
      title={Band-Like Transport and Cation Off-Centring in Ag/Bi-Based Solar Absorbers}, 
      author={Yi-Teng Huang and Yixin Wang and Georgia Fields and Peixi Cong and Yongjie Wang and Jack E. N. Swallow and Avari Roy and Jack M. Woolley and Victoria Rotaru and Maxim Guc and Lars van Turnhout and Mohamed Aouane and Emmanuelle Suard and Dominik Kubicki and Alejandro Pérez-Rodríguez and Aditya Sadhanala and Akshay Rao and Dennis Friedrich and Robert S. Weatherup and Simon J. Clarke and Seán R. Kavanagh and Robert L. Z. Hoye},
      year={2026},
      eprint={2602.22024},
      archivePrefix={arXiv},
      primaryClass={cond-mat.mtrl-sci},
      url={https://arxiv.org/abs/2602.22024}, 
}

@article{2010-ML-AgBiS2-xrd,
title = {Characterization of AgBiS2 nanostructured flowers produced by solvothermal reaction},
journal = {Mater. Lett.},
volume = {64},
number = {6},
pages = {755-758},
year = {2010},
issn = {0167-577X},
doi = {https://doi.org/10.1016/j.matlet.2010.01.003},
url = {https://www.sciencedirect.com/science/article/pii/S0167577X10000078},
author = {Titipun Thongtem and Narongrit Tipcompor and Somchai Thongtem},
}

@article{2022-CPB-DFT,
doi = {10.1088/1674-1056/ac89d7},
url = {https://doi.org/10.1088/1674-1056/ac89d7},
year = {2022},
month = {oct},
publisher = {Chinese Physical Society and IOP Publishing Ltd},
volume = {31},
number = {10},
pages = {107105},
author = {Kang, Jun and Zhang, Xie and Wei, Su-Huai},
title = {Advances and challenges in DFT-based energy materials design},
journal = {Chin. Phys. B},
}

@article{2019-PSSA-amorphous,
author = {Ide, Keisuke and Nomura, Kenji and Hosono, Hideo and Kamiya, Toshio},
title = {Electronic Defects in Amorphous Oxide Semiconductors: A Review},
journal = {physica status solidi (a)},
volume = {216},
number = {5},
pages = {1800372},
doi = {https://doi.org/10.1002/pssa.201800372},
year = {2019}
}

@book{2022-amorphous-book,
  title={Amorphous Oxide Semiconductors: IGZO and Related Materials for Display and Memory},
  author={Hosono, Hideo and Kumomi, Hideya},
  year={2022},
  publisher={John Wiley \& Sons}
}

\end{document}